# Extraterrestrial sedimentary rocks on Earth

Yana Anfinogenova,[1]* John Anfinogenov,[2] Larisa Budaeva,[3] and Dmitry Kuznetsov[1]

**Affiliations:**

[1]Yana Anfinogenova, Ph.D., National Research Tomsk Polytechnic University

[2]John Anfinogenov, Tunguska Nature Reserve, Ministry of Natural Resources and Ecology of the Russian Federation

[3]Larisa Budaeva, National Research Tomsk State University

*Corresponding author: Yana Anfinogenova. Address: TPU, 30 Lenin Ave., Tomsk, 634050, Russia. Tel: +79095390220. E-mail: anfiyj@gmail.com and anfy@tpu.ru

**Abstract**

This concept article discusses the possibilities for identifying sedimentary-origin meteorites. The paper concerns (i) the macroscopic candidate for sedimentary meteorite in the epicenter of the 1908 Tunguska catastrophe; (ii) potential parent bodies for sedimentary meteorites; (iii) isotopic heterogeneity of unmixed silicate reservoirs on Mars; (iv) possible terrestrial loss or contamination in the noble gas signatures in new type meteorites that spent time in extreme weather conditions; (v) cosmogenic isotopes and shielding; and (vi) pseudo meteorites. We conclude that the list of candidate parent bodies for sedimentary meteorites includes, but is not limited by the Earth, Mars, Enceladus, Ganymede, Europa, and hypothetical planets that could exist between orbits of Mars and Jupiter in the past. A parent body for extraterrestrial sedimentary rocks on the Earth should be identified based on the entire body of evidence which is not limited solely by tests of oxygen and noble gas isotopes whose signatures may undergo terrestrial contamination and may exhibit significant heterogeneity within the parent bodies. Observed fall of cosmic body, evidence of hypervelocity impact complying with the criteria of impact structures, and the presence of fusion crust on the fragments should be considered as priority signs of meteoritic origin.

**Introduction**

A catastrophic collision of cosmic body with the Earth occurred above the Tunguska region of Siberia on June 30, 1908. About 2000 km$^2$ of taiga forest was devastated by shock waves and fire following the major explosion in the atmosphere. A possible impact crater filled up by a 300 m diameter lake (Lake Cheko) has been reported ca. 8 km NNW of the Tunguska event epicenter [Gasperini *et al*. 2007, 2008]. Studies propose cometary [Gladysheva 2011, 2-13] and asteroid [Sekanina 2008, Chyba *et al*. 1993] origin of the Tunguska projectile. Kvasnytsya *et al*. [2013] believe it was an iron meteorite though no sizable fragments of iron meteorite have been recovered throughout the region of the catastrophe.

John Anfinogenov proposed a hypothesis for the existence of sedimentary-origin meteorites (hereinafter referred to as "sedimentary meteorites") after he found a fresh impact site in the epicenter of the 1908 Tunguska explosion containing the exotic sedimentary boulder (John's stone or John's rock) whose splinters had glassy surface reminiscent of freshly applied enamel or fusion crust. His reports on discovery of John's rock and hypothesis of sedimentary meteorites from Mars or from hypothetical planet called Phaeton [McSween 1999] were published by local mass media in 1973 [Anfinogenov 1973]. Findings were further discussed in more detail [Anfinogenov *et al.* 1998b] and presented to international audience [Anfinogenov *et al.* 1998a, 2014]. John's rock is the first macroscopic candidate for a fragment of the Tunguska projectile and for a new type sedimentary meteorite composed of silica-rich metamorphic rock.

Inferred extraterrestrial origin of the Tunguska boulder, discovered by Anfinogenov *et al*. [2014], is compatible with the presence of hydrothermal silica-rich deposits on Mars [McLennan 2003, Bandfield *et al*. 2004, Milliken *et al*. 2008, Squyres *et al*. 2008, Smith *et al*. 2012] as well as with the presence of liquid water [Heller *et al*. 2015, Saur *et al*. 2015, Steinbrügge *et al*. 2015, Tanigawa *et al.* 2014, Vance *et al.* 2014] and hydrothermal activity [Hsu *et al.* 2015, Postberg *et al.* 2011] on several other bodies of the Solar System such as icy moons of Jupiter and Saturn. Mars is not the only candidate parent body for sedimentary meteorites. We hypothesize that sedimentary rocks may form in the presence of water flows generated by tidal and volcanic forces in the oceans on the satellites of the giant planets.

In the present paper, we discuss potential parent bodies for sedimentary meteorites; isotopic heterogeneity of unmixed silicate reservoirs on Mars; possible terrestrial loss or contamination in the noble gas signatures in meteorites that spent time in the extreme weather conditions; and cosmogenic isotopes and shielding hampering identification of new type meteorites. We emphasize the significance of the first macroscopic evidence for a candidate meteorite of a new type (sedimentary-origin meteorite) from Tunguska.

Sedimentary meteorites, found on Earth, will contribute significantly to elucidating the history of Solar System objects and to the search for possible life forms.

**"If a meteorite of Martian sandstone hit you on the head would you recognize it?" [Ashley *et al.* 1999]**

Experts in sedimentary geology Ashley G. M. and Delaney J. S. [1999] believe that sedimentary meteorites should be sampled on Earth in proportion to the fraction of Mars covered by sedimentary rock. They present compositions of Barnacle Bill and Yogi from Sagan. Sta., Mars on classification diagram for igneous rocks with composition ranges of typical sedimentary rocks superimposed. According to the diagram, SNC meteorites, the only rocks generally accepted to form on the planet Mars, represent only a small portion of what types of meteorites from this planet might be found [Ashley *et al.* 1999]. Ashley G. M. and Delaney J. S. [1999] emphasize that fusion crust is crucial for recognition of extraterrestrial origin of meteorites and state that "if a consolidated siliciclastic sediment were ejected from Mars, the fusion crust formed during its deceleration and descent to Earth could be quite unlike anything that previous meteoritic experience defines as true fusion crust" [Ashley *et al.* 1999].

To study the formation of fusion crust on the Martian analogue sediments, a series of experiments determined the effects of thermal alteration of samples during atmospheric entry [Foucher *et al.* 2009, Brack *et al.* 2002]. In these experiments, the holders with terrestrial sedimentary rock specimens were attached to the capsule near the stagnation point of FOTON, robotic spacecraft used by Russia and the European Space Agency. Data show that the admixture of fragments of silicified volcanic sediments and space cement survive the thermal shock well, forming a white fusion crust [Foucher *et al.* 2009].

Available literature contains a few reports on candidate sedimentary meteorites with observed falls. Cross F. C. reported on three rocks found in 1947 including two grayish fine-grained sandstone specimens found in the United States that, in his opinion, deserved consideration to be of cosmic origin [Cross 1947]. Cross F. C. reviewed data of Dr. Assar Hadding, the Director of the Geological Institute in Lund (Sweden), who described two specimens, one of limestone and one of sandstone, that he believed were meteorites [Hadding 1940]. An unusual quartz pebble shower occurred during snowstorm in Trélex, Switzerland on February 20, 1907 [Rollier 1907]. Pebble sizes ranged from pea to hazelnut. These meteor pebbles were composed of milky quartz and were not included in large hail. The origin of pebbles was not identified with certainty. It remains unclear whether these pebbles were meteorites or they could come from the Mediterranean Sea (Islands of Hyères) or the Meseta (Spain) which is even farther away from Trélex.

We found a total of 226 records for meteorites with status of "pseudo" in the category of "terrestrial meteorites" in the Meteoritical Bulletin Database of the International Society for Meteorites and Planetary Science [The Meteoritical Society 2016]. Except NWA 6944, all other "pseudo" meteorites were found in Antarctica. No characteristics are provided for description of these rocks, but the fact that these specimens were considered candidates for meteorites and were included in the Meteoritical Bulletin Database might suggest that their minerology was exotic to the areas of discovery and/or they presented with a glassy cover or a fusion crust-like surface. It would be helpful to know more about those rocks. Sedimentary meteorites might be present among them.

**John's rock from the epicenter of the 1908 Tunguska catastrophe**

John's rock is an exotic boulder (Fig. 1) found on the Stoykovich Mountain in the epicenter area of the 1908 Tunguska catastrophe [Anfinogenov *et al*. 2014]. Pattern of permafrost destruction suggests high-speed entry and lateral ricochet of John's rock with further deceleration and breakage producing the ~50-$m^3$ impact groove in the permafrost (Fig. 2). Landing velocity of John's rock is estimated to be at least 547 *m*/s [Anfinogenov *et al*. 2014]. John's rock is composed of highly silicified gravelite sandstone (~99% $SiO_2$). Outer surface of several splinters of this rock shows continuous glassy coating reminiscent of fresh enamel or fusion crust. There is a clear consistency in geometry of meteoroid flight trajectory, locations of John's rock fragments and cleaved pebbles, and directions of impact groves left by the largest fragments (Fig. 2). John's rock locates on quaternary deposits at the top of the Stoykovich Mountain and is exotic to the territory within at least hundreds of kilometers around the epicenter [Sapronov 1986]. This rock significantly differs from local tufogenic sandstones, the fact which has been discussed in detail before [Anfinogenov *et al.* 2014, Sapronov 1986]. There are no signs of past glaciation throughout the region of the 1908 Tunguska catastrophe [Sapronov 1986]. Decoding of aerial survey photographs covering area within 40 km from the epicenter shows the absence of active diatremes [Anfinogenov *et al.* 2014].

John's rock did not produce a typical impact crater on the Stoykovich Mountain in the 1908 Tunguska explosion epicenter. Instead it formed the massive (at least 50 m$^3$ in volume) canal and the groove in the permafrost (Fig. 2). The absence of classic impact crater is similar to several other cases when large meteorites fall on hydric soils, sands, or mountain shoulders sloping down along the motion of cosmic body producing atypical impact formations. For example, among thousands of fragments of the 1947 Sikhote-Alin meteorite, several large fragments fell on frozen hydric soil forming up to 8-m long canals with small entry holes (Fig. 3) [Krinov *et al.* 1959]. Similarly, the largest fragment of the Chelyabinsk meteorite, a ~570 kg rock, did not form a significant impact crater at 20 m deep bottom of Lake Chebarkul despite high impact velocity for this fragment was estimated to be 19 km/s [Chapman 2013, Popova 2013]. We find this value of velocity overestimated as it corresponds to kinetic energy of 25 tons of TNT which would be significantly more destructive. Perhaps, both the largest Chebarkul fragment of the Chelyabinsk meteorite and John's rock in Tunguska landed at low supersonic speeds without producing major impact craters.

The phenomenon of John's rock largely complies with the criteria of impact structures [French and Koeberl 2010]. There are melted, shocked, and brecciated rocks associated with this boulder [Anfinogenov et al. 2014]; some of them fill the resulting disruption in permafrost [Anfinogenov et al. 2014], and others are transported to some distances from the source impact funnel [Anfinogenov et al. 2014]. There is the presence of diagnostic shockmetamorphic effects in John's rock [Anfinogenov et al. 2014, Bonatti et al. 2015]. Shock deformation associated is expressed in macroscopic form [Anfinogenov et al. 2014] and in microscopic forms (e.g., deformation lamellae in quartz) [Bonatti et al. 2015]. There are macroscopic data and mathematical calculations showing that John's rock is associated with hypervelocity impact [Anfinogenov et al. 2014].

There is a significantly higher (by hundred-fold) content of glassy silicate microspherules precipitated from the atmosphere in the peat layer of 1908 [Dolgov *et al.* 1973] throughout the entire region of the 1908 Tunguska catastrophe. Composition of these microspherules is consistent with material of John's rock. Together with macroscopic evidence of John's rock impact, data on glassy silicate microspherules suggest that John's rock may be a fragment of the 1908 Tunguska meteorite and may represent a new type of meteorites of planetary origin [Anfinogenov *et al.* 2014].

Bonatti *et al.* [2015] reports on structure, mineralogy and chemistry of John's rock suggesting that it may have originated by silica deposition from hydrothermal solutions that reacted with basaltic rocks. Oxygen isotope data suggest that the precipitation of $SiO_2$ could have occurred in equilibrium with hydrothermal water ($\delta^{18}O_w \approx -16$ ‰) at the temperature of about 80°C [Bonatti *et al.* 2015]. High precision triple oxygen isotope data reveal that this rock is inconsistent with the composition of known Martian meteorites [Haack *et al.* 2015, Bonatti *et al.* 2015]. Notably, the Tunguska boulder, in addition to $SiO_2$, contains traces of a Ti-oxide phase and its bulk composition [Bonatti *et al.* 2015] is similar to the composition inferred from APXS data [Squyres *et al.* 2008] for the silica deposits of the Gusev crater on Mars. A paper by Bonatti *et al.* [Bonatti *et al*. 2015] contains a section with detailed review of sedimentary rocks and hydrothermal activity on Mars.

Instrumental data cannot unambiguously favor terrestrial or extraterrestrial origin of John's rock. There are pros and contras [Bonatti *et al.* 2015] of extraterrestrial origin of this boulder. However, the presence of fusion crust-like surface on the fragments and signs of high-speed impact associated with John's rock in the epicenter of the 1908 Tunguska explosion represent direct evidence for its meteoritic nature [Anfinogenov *et al.* 2014]. Further in-depth study of this rock should be undertaken including the thermoluminescence analysis, rock age determination, and the comparison of John's rock signatures with those of similar terrestrial and extraterrestrial rocks. The site of impact associated with John's rock requires comprehensive interdisciplinary field examination.

**Other geomorphological phenomena associated with the 1908 Tunguska catastrophe**

John's rock is not the only object associated with the 1908 Tunguska catastrophe. Eyewitnesses who personally visited the region of the 1908 Tunguska catastrophe soon after the event reported the presence of several *de novo* geomorphological phenomena in the impact area: fountain-like ejecta of boggy deposits associated with the fall of the Tunguska cosmic body; flowing of water well from under the ground for several days after the event; funnel-like hole in the ground at the end of the *de novo* dry groove conventionally named "dry river"; ground disturbance in the form of a groove with rocks in the walls; and boulders of unusual color and texture that appeared "out of thin air" in the forest after the event.

Decoding of the aerial survey photographs taken in 1938 and 1949 and the field studies performed from 1965 to 2015 confirmed that the fall of the Tunguska cosmic body fragments indeed caused the formation of impact structures such as the funnels, grooves, and the pipe-like disturbances in the ground. The fall also produced powerful surface seismic waves triggering massive rock-slides and disturbances in surface and subsurface hydrogeological structures over the area of the Tunguska catastrophe. The groove- and pipe-like disturbances in the ground, associated with John's rock, represent the impact structures with morphology distinct from that of classical impact craters. In the epicenter area of the 1908 Tunguska catastrophe, other pipe-like impact structures, containing large boulder-sized meteorite rocks, may be present [Anfinogenova *et al.* 2016].

**Could be the Tunguska cosmic body a rubble pile asteroid partially composed of extraterrestrial sedimentary rock?**

The discovery of significantly higher content of glassy silicate microspherules precipitated from the atmosphere in the peat layer of 1908 throughout the area of the Tunguska catastrophe was previously reported [Dolgov *et al.* 1973]. The peat layer of 1908 contains up to hundredfold-higher count of gray and colorless transparent silicate microspherules than the adjacent peat layers. Data of neutron activation analysis show that chemical composition of microspherules was distinct from that of industrial glass, local terrestrial microparticles, known stony meteorites, tektites, and Moon rocks [Kolesnikov *et al*. 1976]. The anomaly of glassy silicate microspherules/quartz grains [Dolgov *et al.* 1973] provides a statistically significant evidence for silica-rich projectile responsible for the 1908 Tunguska event. Even more so, the quartz grains were found in sediment cores collected from Lake Cheko [Gasperini *et al.* 2009] suggesting that they might have resulted from dust produced by the explosion in the atmosphere of the main body if the Tunguska cosmic body were silica-rich. The presence of this silica anomaly is consistent with the hypothesis on extraterrestrial origin of John's rock.

However, anomalous presence of glassy silicate microspherules in the 1908 peat layer is not the only anomaly known in the region of the 1908 Tunguska catastrophe. Rasmussen *et al.* [1999] and Hou *et al.* [2004] reported the presence of other particles suggesting PGE anomaly in peat cores also at an estimated depth of 1908 in Tunguska. It is essential to note that Tunguska explorer Dr. E. M. Kolesnikov who performed the aforementioned neutron activation analysis [Kolesnikov *et al.* 1976] of anomalous silicate microspherules from peat layer of 1908 is also a co-author of both papers by Rasmussen *et al.* [1999] and by Hou *et al.* [2004] reporting PGE anomaly associated with particles of different nature. Therefore, two distinct anomalies are present in the area of interest and they both are reported by the same Tunguska meteorite explorer, Dr. E.M. Kolesnikov: (i) anomalous abundance of glassy silicate microspherules precipitated from the atmosphere in the peat layer of 1908 and (ii) PGE anomaly.

Studies by Rasmussen *et al.* [1999] and by Hou *et al.* [2004] produced data that could shed some light on the nature of the Tunguska meteorite. However the methodology of these studies may be criticized. These studies are based on data from an insignificant number of peat core samples showing PGE anomaly: only four adjacent peat columns and only one peat column were studied by Rasmussen *et al.* [1999] and by Hou *et al.* [2004], respectively. Peat cores from singular topographic locations were tested in these studies. On the contrary, Dolgov *et al.* [1973] described anomalous content of glassy silicate microspherules in the peat layer of 1908 in hundreds of peat samples associated with the entire area of the 1908 Tunguska catastrophe. Nevertheless, we admit the presence of these two anomalies associated with the peat layer of 1908 in Tunguska.

Two explanations may be proposed for the presence of these two anomalies (PGE and glassy silicate microspherules/quartz grains) in the area of the 1908 Tunguska catastrophe. According to the first explanation, two impact events involving silica-rich impactor and chondritic or cometary projectile happened independently within relatively short period of time.

According to the second explanation, the impactor could have a complex conglomerate composition consisting of multiple parts that merged due to either collision of parent asteroids in outer space or due to a co-ejection of different adjacent rocks from a parent planetary body after a high-energy impact. For example, it could be a co-ejection of enclosing bedrocks, intrusive igneous rocks, and impactor material where any of these components could partially melt due to impact on parental planetary body. Such processes could produce a rubble-pile asteroid causing dual anomaly at the impact site in Tunguska.

Interestingly, no macroscopic pieces of chondritic or cometary projectile have ever been found in the area of the 1908 Tunguska catastrophe. Chyba *et al.* [1993] reported in Nature magazine that carbonaceous asteroids and especially comets are unlikely candidates for the Tunguska object. He states that the Tunguska event represents a typical fate for stony asteroids tens of meters in radius entering the Earth's atmosphere at common hypersonic velocities [Chyba *et al.* 1993]. In this regard, John's rock represents a sound macroscopic candidate for a stony impactor though of a previously unknown type. This hypothesis is consistent with John's rock phenomenon bearing the numerous signs of high-speed impact and glassy fusion crust-like surface on some splinters. It is also consistent with the discovery of the quartz grains in sediment cores collected from Lake Cheko [Gasperini *et al.* 2009] and the glassy silicate microspherules anomaly associated with the area of the 1908 Tunguska catastrophe [Dolgov *et al.* 1973, Kolesnikov *et al.* 1976].

Considering the presence of two distinct anomalies, the Tunguska cosmic body may be a rubble pile asteroid partially consisting of a sedimentary rock. Neither the Earth nor other bodies in the Solar System can be ruled out as candidate parent bodies of John's rock as potential sedimentary meteorite based on the presence of PGE anomaly reported by Rasmussen *et al.* [1999] and Hou *et al*. [2004].

**Potential parent bodies**

Haack *et al.* [2015] argue that "since sandstones can only form on a parent body with liquid water and, by inference also an atmosphere… there are only two possible parent bodies in the Solar System: the Earth and Mars". Let us discuss this limitation. In the Solar System, many bodies without a significant atmosphere have abundant liquid water [Heller *et al.* 2015, Saur *et al.* 2015, Steinbrügge *et al.* 2015, Tanigawa *et al.* 2014, Vance *et al.* 2014]. Hydrothermal processes occur, in particular, on Enceladus [Hsu *et al.* 2015, Postberg *et al*. 2011]. The presence of powerful tidal currents of water in subsurface oceans on icy satellites of Saturn and Jupiter may provide conditions necessary and sufficient for formation of sedimentary and metamorphic rocks including sandstones and pebble conglomerates with a variety of grain sizes.

The Saturnian moon Enceladus may be considered a candidate parent body for sedimentary meteorites composed of highly metamorphic rock due to (i) the presence of global hydrothermal activity, (ii) the presence of powerful tidal currents of water in subsurface oceans potentially resulting in formation of sediments, and (iii) the past history of large-scale impacts explaining ejection of Enceladus' crust fragments into space. Interestingly, the plume of Enceladus emits nanometre-sized $SiO_2$ (silica)-containing ice grains [Hsu *et al.* 2015] formed as frozen droplets from a liquid water reservoir contacting with silica-rich rock [Postberg *et al.* 2011]. Characteristics of these silica nanoparticles indicate ongoing high temperature (>90 °C) global-scale geothermal and hydrothermal reactions on Enceladus favored by large impacts [Hsu *et al.* 2015]. Enceladus has a differentiated interior consisting of a rocky core, an internal ocean and an icy mantle. Simulation studies suggest that large heterogeneity in the interior, possibly including significant core topography may be due to collisions with large differentiated impactors with radius ranging between 25 and 100 km. Impacts played the crucial role on the evolution of Enceladus [Monteux *et al.* 2016] and similar effects on evolution are very likely on the other moons of Saturn as well as on other planetary objects, such as Ceres [Davison *et al.* 2015, Ivanov 2015].

A putative subsurface water ocean is present on Ganymede. Iron core of Ganymede is surrounded by a silicate rock mantle and by a globe-encircling, briny subsurface water ocean with alternating layers between high pressure ices and salty liquid water [Saur *et al*. 2015, Steinbrügge *et al.* 2015, Vance *et al.* 2014]. If Ganymede or Callisto had acquired their $H_2O$ from newly accreted planetesimals after the Grand Track [Mosqueira *et al.* 2003], then Io and Europa would be water-rich, too [Heller *et al.* 2015, Tanigawa *et al.* 2014].

Tidal dissipation and tidal resonance in icy moons with subsurface oceans are major heat sources for these icy satellites of the giant planets [Kamata *et al.* 2015]. Tidal forces generate heat and currents of liquid water or brine powerful enough to produce sediments that undergo metamorphic transformations due to hydrothermal activity.

Therefore, candidate parent bodies for sedimentary meteorites are not limited by the Earth and Mars only.

**Novel scenario: Hypothetical planet Phaeton as parent body of some asteroids and meteorites including extraterrestrial sedimentary rocks**

In early 1970s, after discovering sedimentary boulder associated with high-speed impact in the epicenter of the 1908 Tunguska catastrophe, John Anfinogenov [1973] hypothesized that sedimentary meteorites may come to the Earth from hypothetical planet called Phaeton [McSween 1999]. Findings of so-called Martian meteorites and some pseudo-meteorites belonging to upper-crust rocks (volcanic and highly-metamorphic igneous and sedimentary rocks) from Mars-like planets provide rationale for reconsideration of the hypothesis on past existence of a planet between the orbits of Mars and Jupiter as a parent body of the asteroid belt. Nature of the hypothetical planet and mechanism of its total explosive disintegration remain the main questions. Various mechanisms were proposed to answer the challenges of the exploded planet hypothesis [Van Flandern 2007]. Considering interorbital structural characteristics of the Solar system, we believe that two or even three planets could be spaced between the orbits of Mars and Jupiter. Providing certain geometry of their orbits and influence of the giant planet Jupiter, these neighbor planets could close with each other up to catastrophic collision.

Based on mathematical modeling with iterative refining, here we propose a scenario involving two hypothetical planets Phaeton I and Phaeton II with the following characteristics: (i) masses comparable with the mass of Mars; (ii) average distances of 2.4 A.U. between the Sun and Phaeton I and 3.95 A.U. between the Sun and Phaeton II; (iii) elliptic orbits like those of Pluto and Mercury with the major axes in the ecliptic plane; (iv) orbital plane inclinations of 15° relative to the ecliptic plane, and the angle of 30° between the orbital planes of these planets; (v) similar Phaeton II's perihelion and Phaeton I's aphelion distances (2.9 ± 0.1 A.U. from these planets to the Sun, respectively) and the distance between these planets of ≤0.1 A.U. at their closest approach to each other.

The catastrophic collision was possible in case of spatial and temporal co-occurrence of Phaeton II's perihelion and Phaeton I's aphelion if the courses of these planets were intersecting at 30° angle; if individual orbital velocities of Phaeton I and Phaeton II were ~16 and ~20 km/s, respectively; and if their closing velocity was ~9 km/s at the moment of the collision. If this was the case, different variants of a spacial arrangement of Phaeton I and Phaeton II physical bodies relative to each other were possible including a scenario where Phaeton I collided with a rear hemisphere of Phaeton II (billiard ball impact type). If closing velocity was ~9 km/s, the impulse of impact force and the kinetic energy were sufficient for disintegration of both planets and dispersion of their fragments with the velocities exceeding second cosmic speed for their masses. In such a case, a significant portion of the mass from Phaeton II and its fragments might acquire an additional orbital acceleration accounting for the stretching of their orbit up to the Jupiter orbit with a possible capture of some Phaeton II fragments by Jupiter so these fragments assumed orbital motion around Jupiter and even the entry into the Jovian atmosphere. A significant portion of the mass from Phaeton I might undergo deceleration which caused the exit of its fragments from orbit and the acceleration towards the Sun and the Earth-type planets with a possible fall on them. Middle part of the zone where two planets collided could form the asteroid belt.

We believe that the layered structure of Mars' moon Phobos may suggest its origination from the planetary crust of Phaeton I. This notion agrees well with work of Simioni E. *et al.*

[2015] who proposes the explanation of the observed distribution of the grooves on Phobos as remnant features of an ancient parent body from which Phobos could have originated after a catastrophic impact event [Simioni E. *et al.* 2015]. We hypothesize that Olympus Mons, the largest volcano in the Solar system, could form when the other large fragment of Phaeton I collided with Mars and produced gigantic impact hole in its planetary crust.

Proposed reconstruction of the initial planetary structure of the Solar system and its partial disruption are compatible with the structural and dynamic characteristics of the preserved parts of the Solar system. Identification of progenitors for the meteorites and the bodies from the asteroid belt as well as determination of cosmic age of their formation require to consider that some of them may belong to the rocks originating from planetary crusts of hypothetical planets Phaeton I and Phaeton II.

Therefore, Hypothetical planets Phaeton I and Phaeton II that possibly existed in the past may represent parental bodies of putative new-type planetary meteorites including Martian-like meteorites and meteorites belonging to upper-crust rocks such as volcanic and highly-metamorphic igneous and sedimentary rocks.

**Isotope tests of candidate extraterrestrial sedimentary rock from the 1908 Tunguska explosion epicenter**

Isotopic characterizations of the exotic boulder from the epicenter of the 1908 Tunguska event are essential and further elemental and isotopic characterizations of this rock are highly encouraged. High precision triple oxygen isotope data reveal that this rock is inconsistent with the composition of known Martian meteorites [Haack *et al*. 2015, Bonatti *et al*. 2015]. However, numerical results of isotopic characterizations, though important on their own, cannot prove or disprove the origin of the rock. Indeed, no rock samples have ever been delivered from Mars or from other cosmic bodies with abundant water to the Earth before. Isotopic compositions of Martian sedimentary rocks have never been tested by Mars rovers and remain unknown. All which was tested up to day were the Martian meteorites. Meanwhile, diversities in the rock-forming processes and in the corresponding rock types within planets are insufficiently studied. A significant heterogeneity in $\Delta^{17}O$ in rocks of different types has been reported [Wang *et al.* 2013]. Pack and Herwartz [Pack *et al*. 2014, 2015] provide evidence that the concept of a single terrestrial mass fractionation is invalid on small-scale. They conclude that mineral assemblages in rocks fall on individual "rock" mass fractionation lines with individual slopes and intercepts.

We believe the same may be true for other bodies of the Solar System such as Mars, large moons of Jupiter and Saturn, and large asteroids. An idea of separate long-lived silicate reservoirs on Mars is supported by radiogenic isotope studies [Borg *et al*. 1997, 2003]. The distinct $\Delta^{17}O$ and $\delta^{18}O$ values of the silicate fraction of NWA 7034 compared to other SNC meteorites support the idea of distinct lithospheric reservoirs on Mars that have remained unmixed throughout Martian history [Agee *et al.* 2013, Ziegler *et al.* 2013]. Isotopic heterogeneity, including that in the noble gases, can be significant in the Martian mantle. Models for accretion and early differentiation of Mars were tested with chronometers several of which provided evidence of very early isotopic heterogeneity preserved within Mars [Halliday *et al.* 2001]. If significant heterogeneities are reported for known Martian meteorites which are all igneous rocks, then even greater isotopic heterogeneities may exist for the rocks of different types within planet.

The future studies should provide comparative characterizations of particular terrestrial and extraterrestrial rocks similar to John's rock (highly metamorphic highly silicified gravelite sandstones with $SiO_2$ content of nearly 99%), which would be helpful for methodologically proper. Meanwhile, isotopic compositions of John's rock have been compared [Haack *et al.* 2015] only with published data on the Chelyabinsk meteorite [Nishiizumi *et al.* 2013, Busemann *et al.* 2014], SNC meteorites [Franchi *et al.* 1999], NWA 7034 [Agee *et al.* 2013], CIs [Clayton *et al.* 1999], Lunar rocks (no reference was provided by [Haack *et al.* 2015]), angrites and HEDs [Greenwood *et al.* 2005], aubrites and enstatite chondrites [Newton *et al.* 2000], and averaged characteristics of unidentified quartz-rich terrestrial rocks. This comparison is valuable on its own, but it does not prove or disprove extraterrestrial origin of John's rock considering the unique occurrence of this Tunguska boulder.

Tests, performed with the samples of John's rock, should be done with similar extraterrestrial rocks. Such a study may be feasible in the mid-term future considering that matching sedimentary silica-rich rocks have been found on Mars [McLennan 2003, Bandfield *et al.* 2004, Squyres *et al.* 2008, Smith *et al.* 2012, Bandfield 2006, Christensen *et al.* 2005, Edgett *et al.* 2000, Jerolmack 2013, Kerber *et al.* 2012, Michalski *et al.* 2013, Williams *et al.* 2013]. Indeed, there are sedimentary rocks on Mars including pebble conglomerates and sandstones [Kerber *et al.* 2012, Williams *et al.* 2013]. Quartz-bearing deposits are consistently co-located with hydrated silica on Mars [Smith *et al.* 2012]. There is striking visual similarity between Martian pebble conglomerates [Daily Mail Reporter 2013] and the Tunguska boulder associated with the impact [Anfinogenov *et al.* 2014]. Terrestrial rocks reminiscent of John's rock would represent necessary terrestrial control required to be tested by using the same equipment in the same set of experiments. The entire pool of evidence suggesting high-speed impact associated with the John's rock [Anfinogenov *et al.* 2014] should be considered.

Concerning isotopic compositions of John's rock, we also would like to point out here some inconsistencies in data presented by Haack *et al.* [2015]. In particular, Table 1 from [Haack *et al.* 2015] shows that $^{36}$Ar was not detected in four out of five samples of John's rock. Perhaps, inability to detect $^{36}$Ar was caused by its extremely low content. This lack of $^{36}$Ar would agree with data generated by Mars rover suggesting a depletion of Martian atmosphere in light Ar isotopes [Atreya *et al.* 2013] and potential martian origin of John's rock. However, authors [Haack *et al.* 2015] state that the ratio of light to heavy argon isotopes in the samples of John's rock corresponds to the terrestrial one. This statement is highly controversial. How did authors calculate five ratios of $^{36}$Ar/$^{40}$Ar if $^{36}$Ar was not detected in four out of these five samples? We question the interpretation of $^{38}$Ar contents presented in Table 1 by Haack *et al.* [Haack *et al.* 2015] for the same reason.

**Possible terrestrial loss or contamination in the noble gas signatures**

Terrestrial loss or contamination in the noble gas signatures should be considered for any meteorite including Martian meteorites that spent time in terrestrial environment [Schwenzer *et al.* 2013]. All the more so for the Tunguska meteorite because weather conditions in Tunguska are extreme and temperatures range from –61 °C during winter to +40 °C during summer. Considering significant atmospheric precipitations in Tunguska, these conditions are harsher much than those in Antarctica or in deserts. If John's rock is a fragment of the Tunguska impactor, then the repeated freezing of water easily penetrating into tiny pores in its material for decades could significantly accelerate contamination of the rock with terrestrial elements concealing original isotopic signatures.

**Shielding**

The lack of any cosmogenic noble gases such as $^3$He, $^{21}$Ne, $^{38}$Ar would be consistent with an extraterrestrial origin of John's rock under large shielding. We agree with this notion. Indeed, a pre-atmospheric diameter of Tunguska cosmic body is estimated to be tens of meters in diameter [Chyba *et al.* 1993] suggesting up to 50 m of shielding whereas a core part of the Tunguska impactor would have better chances of surviving passage through the atmosphere.

**Summary**

We conclude that the list of candidate parent bodies for hypothetical sedimentary meteorites includes, but is not limited by the Earth, Mars, Enceladus, Ganymede, and Europa. John's rock from the 1908 Tunguska catastrophe epicenter is a candidate sedimentary meteorite composed of silica-rich sedimentary rock. Sedimentary meteorite parent body should be identified based on consensus of all evidence, not limited solely by tests for oxygen and noble gas isotopes which may undergo terrestrial contamination and exhibit significant heterogeneity within the parent bodies. Observed fall of cosmic body, evidence of high-speed impact, and the presence of fusion crust on the fragments should be considered as priority signs of meteoritic origin. We invite scientific community to consider the significance of evidence for the extraterrestrial origin of the exotic sedimentary boulder discovered with signs of recent high-speed impact in the epicenter of the 1908 Tunguska catastrophe. Carl Sagan said once that "extraordinary claims require extraordinary evidence". The macroscopic signs of high-speed impact associated with John's rock, fusion crust-like glassy cover on its fragments, location of this boulder in the epicenter of the 1908 Tunguska catastrophe, and its minerology alien to the region represent extraordinary evidence for extraterrestrial origin of this sedimentary rock.

**Disclosure**

Authors declare that they do not have any financial conflict of interests associated with this article.

**Figure legends**

Figure 1. Photo of John's rock (made by Y. Anfinogenova in 2015).

Figure 2. A: Scheme of John's rock (JR) and its largest fragments as they were found at the Stoykovich Mountain in 1970 and of their reconstruction as before breakage. Bar size = 10 *m*. Fig. B: Side and top views of the high-speed entry, ricochet, and breakage of John's rock whose deceleration in permafrost produced the groove ca. 50 m$^3$ in volume. Ricochet was caused by bouncing of John's rock from dense geological deposits.

Figure 3. An example of the absence of typical impact crater when meteorite fell on wet sandy ground. Only a small funnel (size of 2.3 *m* x 1.8 *m* and depth of 0.9 *m*) was produced by this fragment (255.6 kg) of the Sikhote-Alin iron meteorite fallen in 1947. The fragment fell on wet ground composed of soil (0.2 m), clay (1.8 m), and then sand mixed with red clay. An intact individual exemplar of meteorite was recovered from the bottom of the channel at depth of 6 *m* (Krinov & Fonton, 1959).

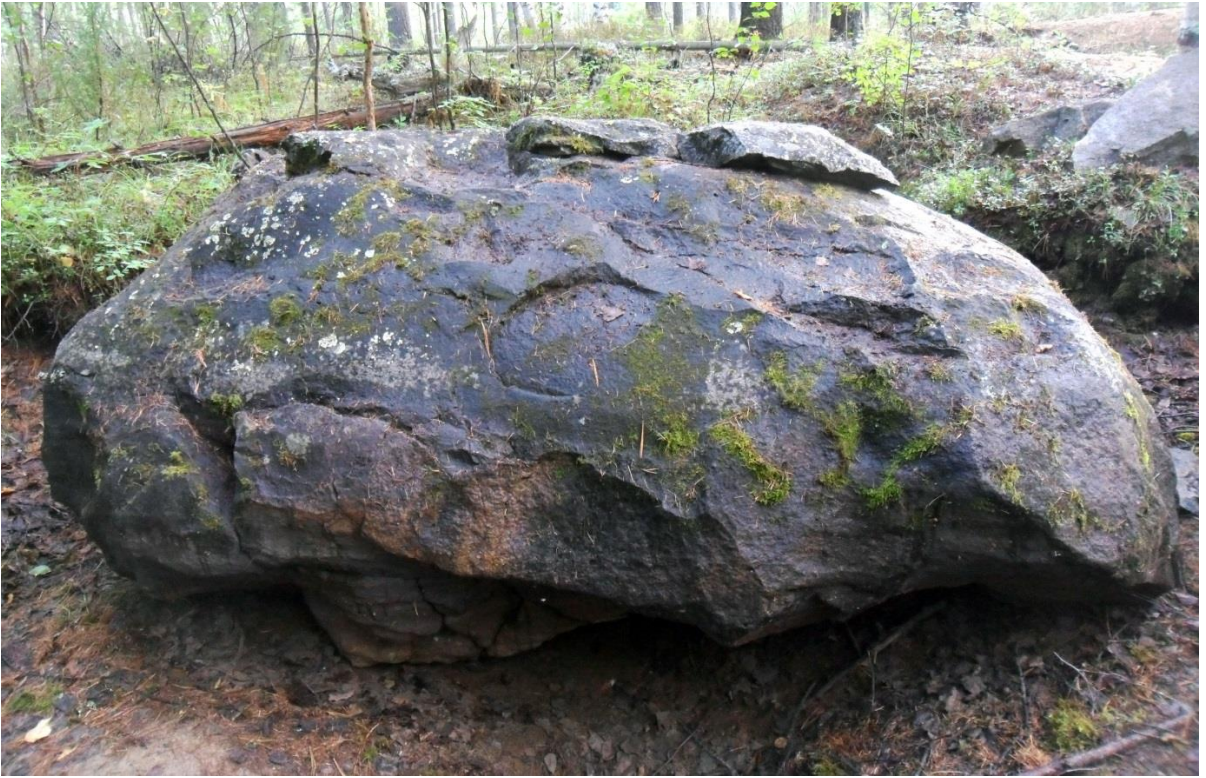

Figure 1.

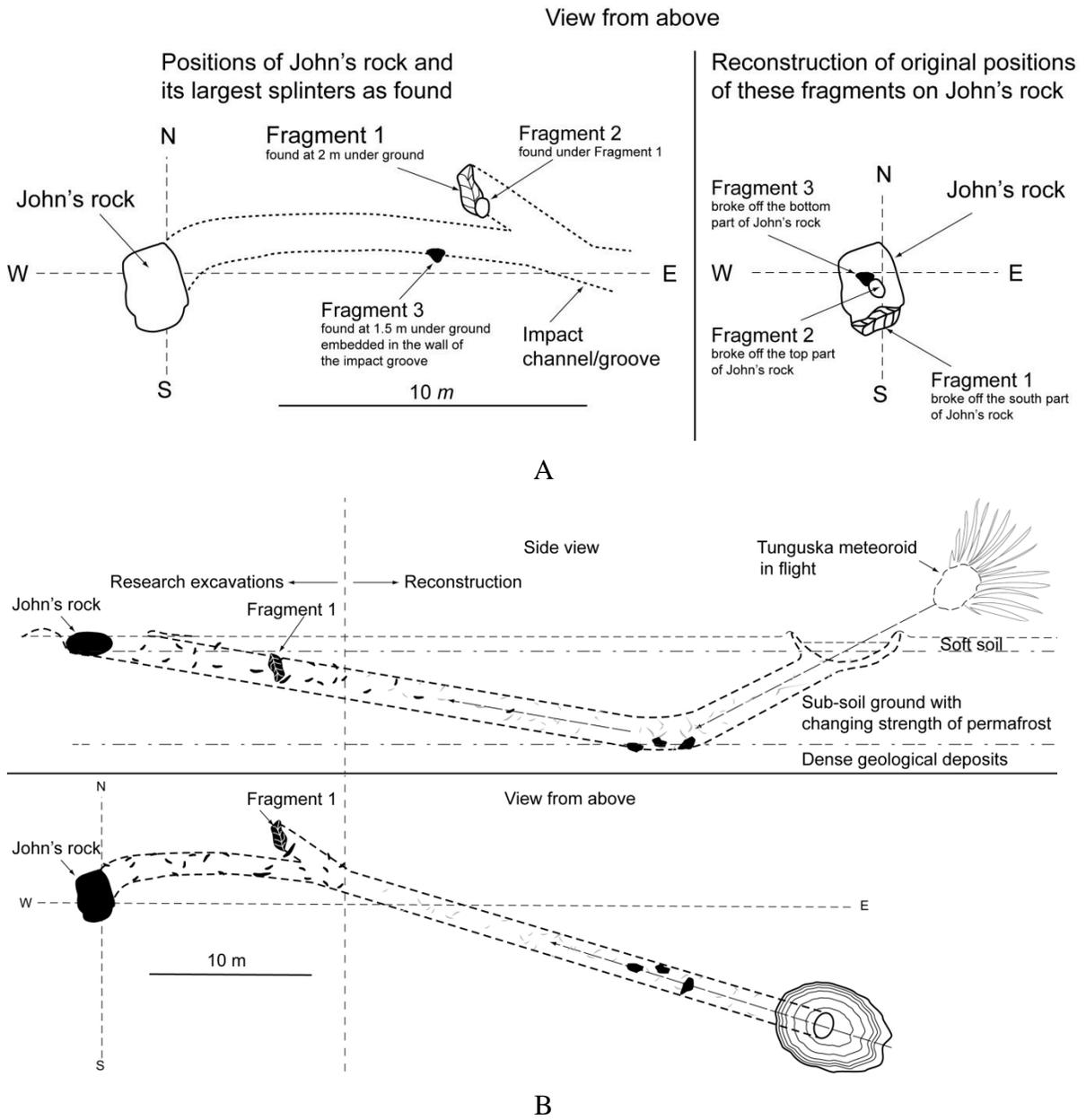

Figure 2.

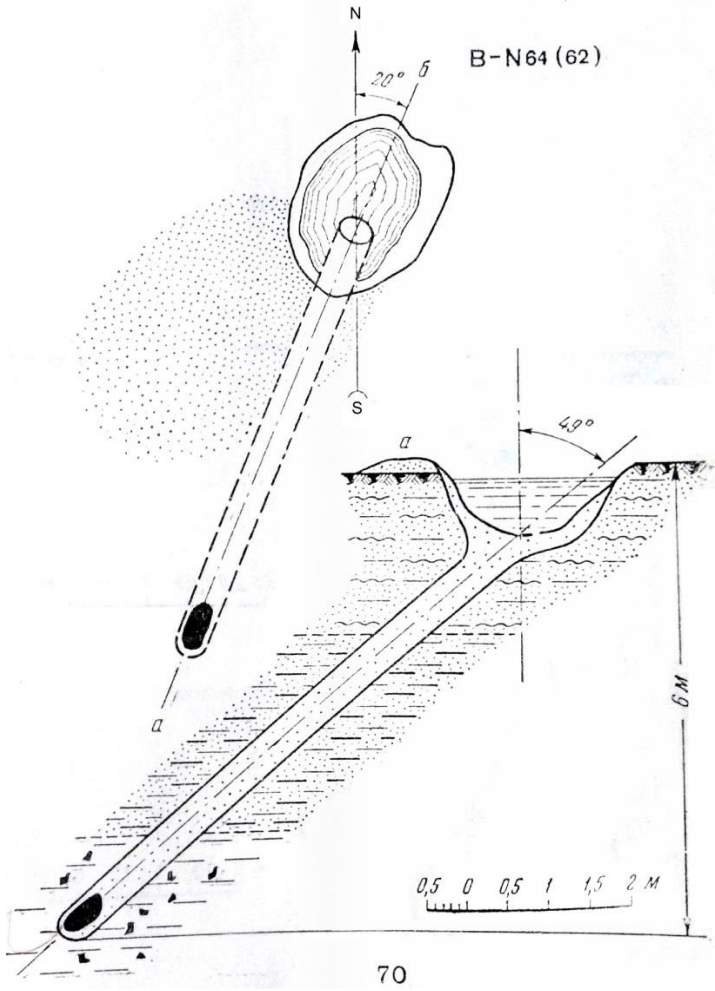

Figure 3.